\documentclass[twoside]{article}
\usepackage{fleqn,espcrc2}
\usepackage{epsfig}
\def\lesssim{\buildrel < \over {_{\sim}}} 
\def\gtrsim{\buildrel > \over {_{\sim}}}
 
\title{Status of $3\nu$ and $4\nu$ scenarios}

\author{  Eligio Lisi
\address{ INFN (Istituto Nazionale di Fisica Nucleare), Sezione di Bari\\
	Via Amendola 173, 70126 Bari, Italy.\ \ 
        E-mail: \tt eligio.lisi@ba.infn.it}}%
\begin{document}

\begin{abstract}
Some aspects of the current neutrino oscillation phenomenology are briefly
discussed, with emphasis on the status of mass-mixing parameters relevant for
scenarios with three active neutrinos $(3\nu)$, and with an additional fourth
sterile neutrino $(4\nu)$.
\end{abstract} \maketitle

\section{Introduction}
\vspace*{-1mm}

At present, the best evidence in favor of $\nu$ oscillations comes from the
Super-Kamiokande (SK) atmospheric $\nu$ experiment  \cite{Kaji},  corroborated
the MACRO and Soudan2 measurements \cite{Rong}.  The SK data, in general,
indicate that (at least) two mass eigenstates (say, $\nu_j$ and $\nu_i$) are
nondegenerate  $(\Delta_{ji}\equiv|m^2(\nu_j)-m^2(\nu_i)|> 0)$, and that
$\nu_j$ is mixed with active flavor states ($\nu_{e,\mu,\tau}$)  and, possibly,
with a fourth sterile state ($\nu_s$),
\begin{equation}
\nu_j=A_e\,\nu_e+A_\mu\,\nu_\mu+A_\tau\nu_\tau+A_s\,\nu_s\ ,
\end{equation}
where one has $|A_\mu|^2\sim O(1/2)$ from the data and
$|A_e|^2+|A_\mu|^2+|A_\tau|^2+|A_s|^2=1$ from unitarity.

I briefly review some mixing scenarios relevant for atmospheric $\nu$
phenomenology, pointing out also their interplay with solar $\nu$'s \cite{Suzu}
and with LSND  \cite{Spen}, as well as some outstanding issues. Such scenarios
are parametrized in Table~1, where $\psi$, $\phi$, and $\xi$ are appropriate
neutrino mixing angles $(s=\sin,\,c=\cos)$, and $m^2$ is the relevant
atmospheric $\nu$ squared mass difference ($m^2=\Delta_{ji}$) in the various
cases (see \cite{Fo96,Fo99,Fo00} for details on the $2\nu$, $3\nu$, and $4\nu$
notation).

Assuming $2\nu$ mixing, the present atmospheric data \cite{Kaji} provide us
with two solid bounds, 
\vspace*{-1mm}
\begin{equation}
2\nu\;({\rm atm.})\;:
\left\{\begin{array}{rcc}
\log_{10} \displaystyle\frac{m^2}{{\rm eV}^2} & \simeq & -2.5 \pm 0.3\ , 
\\[2.9mm]
\sin^2\psi & \simeq & 0.50\pm 0.17\ , \\
\end{array}\right.
\end{equation}
\vspace*{-1mm}
that are not significantly modified  \cite{Fo99,Fo00} when passing to $3\nu$
and $4\nu$ scenarios.%
\footnote{It is more appropriate to quote $\sin^2\mathbf{\theta}$ (or
$\tan^2\mathbf{\theta}$) rather than  $\sin^2 \mathbf{2\theta}$, especially in
contexts where physics can be asymmetric in the two octants of a mixing angle
$\theta$ \protect\cite{Fo96,Fo99}.}

\begin{table}[t]
\caption{\small Atmospheric $\nu$ mixing cases considered in this talk.}
\begin{tabular}{ccccccc}
\hline\\[-3.7mm]
\hline\\[-2.2mm]
Case	    &$\nu_j$&$m^2$&$|A_e|$ &$|A_\mu|$      &$|A_\tau|$     &$|A_s|$ 
\\[1mm]
\hline
\\[-2.8mm]
$2\nu$&$\nu_2$&$\Delta_{21}$&  ---   &$s_\psi$       &$c_\psi$       &  ---    
\\
$3\nu$&$\nu_3$&$\Delta_{32}$&$s_\phi$&$c_\phi s_\psi$&$c_\phi c_\psi$&  ---
\\
$4\nu$&$\nu_4$&$\Delta_{43}$&  ---   &$s_\psi$       &$c_\xi c_\psi$ &$s_\xi c_\psi$ 
\\[1.5mm]
\hline\\[-3.5mm]
\hline
\end{tabular}
\end{table}
\phantom{.}\vspace*{-9mm}
Outstanding $2\nu$ issues: $(i)$ Can the $m^2$ uncertainty be significantly
reduced before future long-baseline projects \cite{Kaji,Hill}? $(ii)$ Is
$\nu_{\mu,\tau}$ mixing exactly maximal $(\sin^2\psi=1/2)$ or not? $(iii)$
Besides monotonic $\mu$ disappearance, can we  also observe one $\mu$
disappearance+appearance cycle \cite{Anto}?

\vspace*{-.5mm}
\section{$3\nu$ mixing}
\vspace*{-1.5mm}

In atmospheric $\nu$ phenomenology, the small squared mass difference $\delta
m^2 \equiv |m^2_2-m^2_1|$ is often neglected. In this case, $m^2\equiv
|m^2_3-m^2_{1,2}|$ and an extra angle $\phi$ is needed to describe
$(\nu_e,\nu_3)$ mixing (see Table~1). Detailed analyses of SK data show that
the bounds in Eq.~(2) are basically preserved for unconstrained $\phi$
\cite{Fo99},  and that an upper limit can be placed on $\sin^2 \phi$. Such
limit is significantly strengthened by the CHOOZ reactor results:
\begin{eqnarray}
3\nu&:& 	\sin^2\phi\lesssim 0.31 {\rm\ (SK)}\ ,\\
3\nu&:& \sin^2\phi\lesssim 0.04 {\rm\ (SK+CHOOZ)}\ .
\end{eqnarray}
The bounds (2)--(4) are compatible with all typical solutions to the solar
$\nu$ problem. Figure~1 presents the current solutions in the $\nu_{1,2}$
mass-mixing  plane  $(\delta m^2,\tan^2\omega)$, at  fixed values of $\phi$
$(\omega=\theta_{12})$. Unfortunately, there is still an embarrassing
multiplicity of allowed $(\delta m^2,\omega)$ regions \cite{Suzu,Mont}. 
However, such regions typically favor relatively small values of $\phi$, thus
providing a  nontrivial, independent consistency check of the bounds (3,4) in
$3\nu$ scenarios. 

Outstanding $3\nu$ issues:  $(i)$ Can SK atmospheric data reveal
$\sin^2\phi\neq 0$ \cite{Kaji} ?  $(ii)$ If $\phi\neq 0$, how far can we push
the  experimental sensitivity to Earth matter effects \cite{Chiz} and to
sign($\pm m^2$) \cite{Gave,Lind,Mina} ? $(iii)$ Are there subleading effects
(e.g., CP violation \cite{Gave,Lind,Mina})  induced by $\delta m^2$ on
atmospheric $\nu$ oscillations  (or, conversely, by $m^2$ on solar $\nu$
oscillations) ? $(iv)$ Can the  multiplicity of solar $\nu$ solutions be
reduced  in the near future \cite{Suzu,Pala,Nobl,Mero}?

\vspace*{-1mm}
\section{$4\nu$ mixing}
\vspace*{-1.5mm}

Attempts to reconcile all the solar, atmospheric, and LSND  pieces of evidence
for oscillations  require $4\nu$ scenarios with one sterile neutrino $\nu_s$,
and mass spectra of the kind ``2+2'' \cite{Giun} or ``1+3'' \cite{Giun,Pere}.
In the 2+2 case,  the atmospheric $\nu$ mixing can be approximately described
as in the third line of Table~1,  involving the usual $(m^2,\sin^2\psi)$
parameters plus an additional mixing $\sin^2\xi$ \cite{Marr}. For $\sin^2\xi=0$
($=1$), the atmospheric oscillation channel is purely $\nu_\mu\to\nu_\tau$
($\nu_\mu\to\nu_s$); intermediate values of $\sin^2\xi$ correspond instead to
mixed active+sterile oscillations. A recent analysis of SK atmospheric results 
\cite{Fo00,Marr} in the 2+2 scenario shows that the bounds in Eq.~(2) are not
significantly modified, and that there is  room for relatively large values of
$\sin^2\xi$:
\begin{equation}
4\nu:\ \sin^2\xi \lesssim 0.67 {\rm\ (SK\ atm.)}.
\end{equation}
This result shows that, although SK \cite{Kaji} (as well as MACRO  \cite{Rong})
disfavors pure sterile oscillations $(\sin^2\xi=1)$, one could have sizable
sterile mixing besides the standard $(\nu_\mu\to\nu_\tau)$ mixing \cite{Marr}.

Independently, one can study how the usual $(\delta m^2,\omega)$ solar $\nu$
solutions are modified for generic $\sin^2\xi$ in $4\nu$ models \cite{Giun}. It
turns out that solar $\nu$ data prefer relatively large values of $\sin^2\xi$
\cite{Giun}. One can roughly say that,  at present,
\vspace*{-.5mm}
\begin{equation}
4\nu:\ \sin^2\xi \gtrsim 0.3 {\rm\ (solar)}\ ,
\end{equation}
up to variations related to the chosen solution.

Therefore, unlike the $3\nu$ case, the $4\nu$ case shows some ``tension''
between atmospheric (5) and solar (6) results---not  strong enough, however, to
prevent their combination. From (5,6) one might get a possible indication for
nonzero $\nu_s$ mixing, 
\vspace*{-0.5mm}
\begin{equation}
4\nu:\ \sin^2\xi \sim 0.5\pm 0.2 {\rm\ (solar+atm\ data)}\ ,
\end{equation}
%
which favors maximal amplitude of both active and  sterile channels in
atmospheric and solar $4\nu$ oscillations (``fourfold maximal mixing''
\cite{Fo00}).

Outstanding $4\nu$ issues: $(i)$  $(i)$ Will the solar-atmospheric data
``tension'' be weakened or not by future data? $(ii)$ Can we detect matter
effects associated to $\nu_s$ mixing \cite{Marr} ?  $(iii)$ How can we
discriminate the 2+2 and 1+3 spectra \cite{Giun,Pere} ?

\section{Conclusions} 
\vspace*{-.5mm}

The bounds from atmospheric $2\nu$  oscillations in the $\nu_\mu\to\nu_\tau$
channel [Eq.~(2)]  are robust, and are preserved by $3\nu$ and $4\nu$
extensions. In $3\nu$ scenarios, the SK+CHOOZ data provide a bound [Eq.~(4)]
on the amplitude of  $\nu_e$ mixing. In $4\nu$ scenarios,  the solar+atm data
are still compatible with nonzero  $\nu_s$ mixing  within present
uncertainties [Eq.~(7)].  Further theoretical and experimental work is needed
to  nail down the atmospheric $\nu$ mixing amplitudes in Table~1, as well as
to reduce the multiplicity of solar $\nu$ solutions in Fig.~1.



\newpage
\begin{figure*}[t]
\vspace{-.5truecm}
\hspace{-1.2truecm}
\epsfig{bbllx=0,bblly=0,bburx=500,bbury=750,
width=15truecm,figure=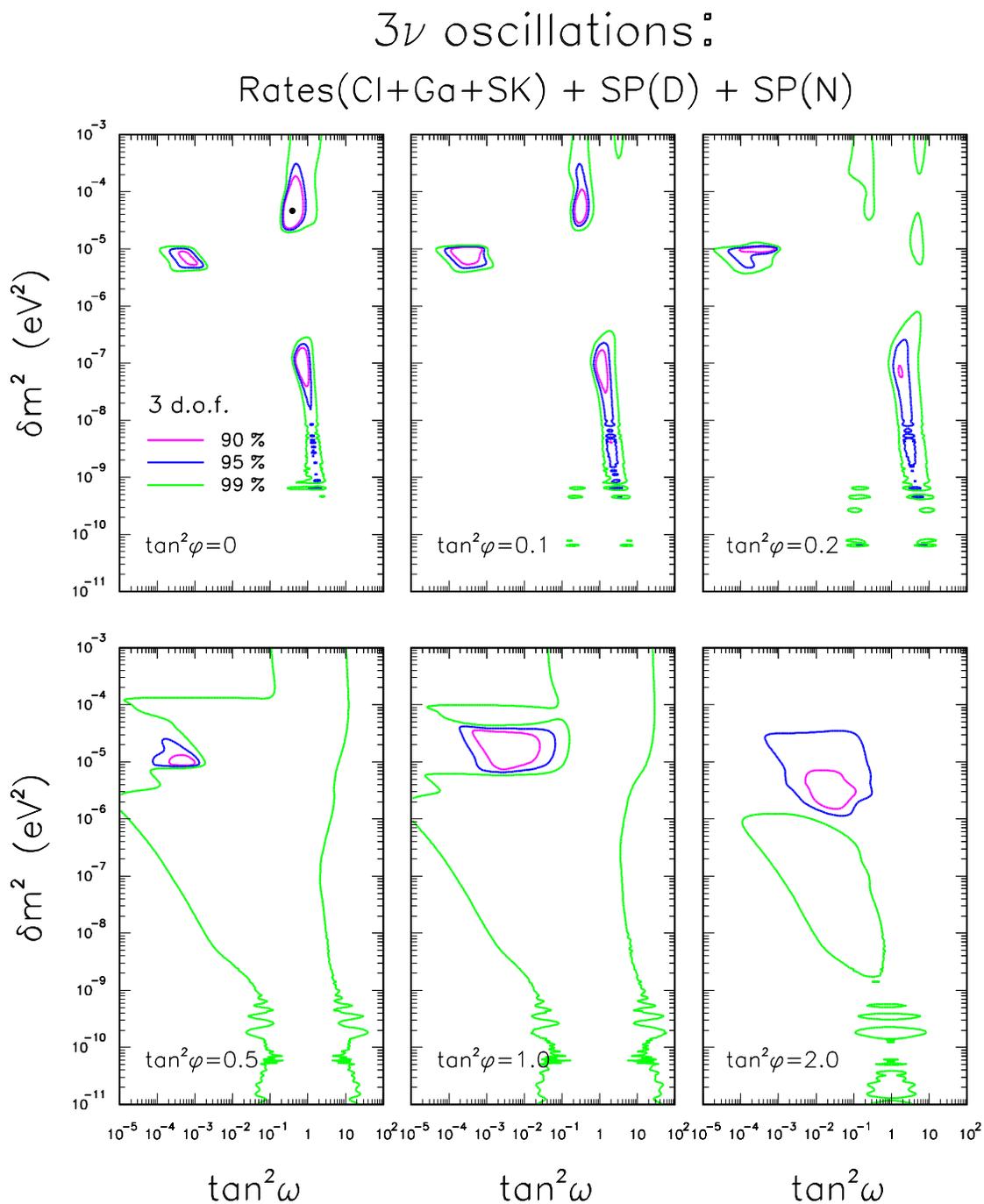}
\vspace{-4.5truecm}
\caption{$3\nu$ solutions to the solar neutrino problem for six representative
values of the angle $\phi=\theta_{13}$. The case $\phi=0$ (first panel)
corresponds to $2\nu$ oscillations. Data: Rates from Homestake,
GALLEX+SAGE+GNO, and SK; day and night  energy spectrum from SK
(updated as of Summer 2000 conferences). Solar Model: BP98 SSM.}
\end{figure*}

\end{document}